# FPGA Acceleration of Short Read Alignment

Nathaniel McVicar, Akina Hoshino, Anna La Torre, Thomas A. Reh,
Walter L. Ruzzo and Scott Hauck

**Abstract**— Aligning millions of short DNA or RNA reads, of 75 to 250 base pairs each, to a reference genome is a significant computation problem in bioinformatics. We present a flexible and fast FPGA-based short read alignment tool. Our aligner makes use of the processing power of FPGAs in conjunction with the greater host memory bandwidth and flexibility of software to improve performance and achieve a high level of configurability. This flexible design supports a variety of reference genome sizes without the performance degradation suffered by other software and FPGA-based aligners. It is also better able to support the features of new alignment algorithms, which frequently crop up in the rapidly evolving field of bioinformatics. We demonstrate these advantages in a case study where we align RNA-Seq data from a hypothesized mouse / human xenograft. In this case study, our aligner provides a speedup of 5.6x over BWA-SW with energy savings of 21%, while also reducing incorrect short read classification by 29%. To demonstrate the flexibility of our system we show that the speedup can be substantially improved while retaining most of the accuracy gains over BWA-SW. The speedup can be increased to 71.3x, while still enjoying a 28% incorrect classification improvement and 52% improvement in unaligned reads.

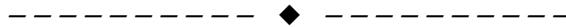

## 1 INTRODUCTION

T HE recent explosion in availability of genome and transcriptome sequencing is revolutionizing medicine and biology. However, it also introduces computational challenges. Chief among these is the processing of the ever-increasing flow of data produced by Next-Generation Sequencing (NGS) machines. High-end modern sequencers can each produce on the order of 100 Gbases of unstructured genetic data per day, making the bioinformatics pipeline one of the more significant big data problems of the current era.

The very first stage of processing this data is often short read alignment. The NGS machine, through chemical processes that vary significantly among different NGS technologies, determines the sequence of numerous short DNA or RNA segments of tens to hundreds of bases. These sequences are known as reads. Alignment attempts to match each read to a known reference genome, such as the human or mouse genome. There are many software packages to perform this alignment, as well as a number of more recent less traditional approaches, including using heterogeneous compute platforms such as systems including FPGAs. Our approach is also FPGA-based, and we make the following important contributions.

— We propose an improved FPGA accelerated short read alignment system. This architecture improves on previous work in terms of performance, flexibility and usability.

— We describe this architecture and the surrounding software system in detail.

— We present a case study comparing this aligner to an earlier FPGA-based system, as well as state of the art software alignment tools. This evaluation demonstrates the performance and accuracy advantages of the system, and introduces a system for aligning short reads against multiple potential reference genomes.

## 2 BACKGROUND

### 2.1 Next-generation Sequencing and Short Reads

During the 1990s the Human Genome Project (HGP) created the first draft sequence of the entire human genome [1][2]. This reference sequence provided a consensus of the nucleotide, or base, for a "typical" human at every position in each of 24 chromosomes. Each DNA base is one of adenine (A), thymine (T), cytosine (C) or guanine (G), so the reference sequence is essentially a string of length ~3.2 billion in a four character alphabet. The first HGP draft sequence required nine years and almost $3 billion to complete due to the relatively immature sequencing technology used [3].

- N. McVicar and S. Hauck are with the Department of Electrical Engineering, University of Washington, 185 Stevens Way, Seattle, WA 98195. E-mail: {nmcvicar, hauck}@uw.edu.

- A. Hoshino and T.A. Reh are with the Department of Biological Structure, University of Washington, 1959 NE Pacific Street, Seattle, WA, 98195. E-mail: {akinah, tomreh}@uw.edu.

- A. La Torre is with the Department of Cell Biology and Human Anatomy, UC Davis School of Medicine, 4303 Tupper Hall, Davis, CA, 95616. E-mail: alatorre@ucdavis.edu.

- W.L. Ruzzo is with the Allen School of Computer Science and Engineering, University of Washington, 185 E Stevens Way NE, Seattle, WA 98195, the Department of Genome Science at the University of Washington and the Fred Hutchinson Cancer Research Center, Seattle, WA, 98109. E-mail: ruzzo@cs.washington.edu.

N. McVicar is the corresponding author.

Next-generation sequencing machines were first introduced in the mid-2000s and have made extremely rapid gains, both in terms of speed and cost. In recent years the decrease in NGS cost has actually outpaced Moore's law [4]. NGS technology has been hugely influential across a number of fields, from archaeology to oncology. While there are a number of different technologies employed in NGS, they are all based on performing many short sequencing operations in parallel. Instead of sequencing an entire chromosome at once (human chromosomes range from ~48 million to ~249 million base pairs [bp]), NGS technology sequences many DNA fragments, ranging from tens to thousands of bp in length, in parallel. These fragments come from replicating the DNA to be sequenced many times and then dividing each copy somewhat randomly, resulting in many overlapping fragments (see Fig. 1). Because of the parallel nature of this technology, a full human genome can currently be sequenced at a cost of roughly $1k to $10k and in a period ranging from a day to a few days. Due to this trend of declining sequencing costs, the bioinformatic analysis required to gain medical or scientific insight from NGS data becomes an ever-increasing percentage of the total time and expense.

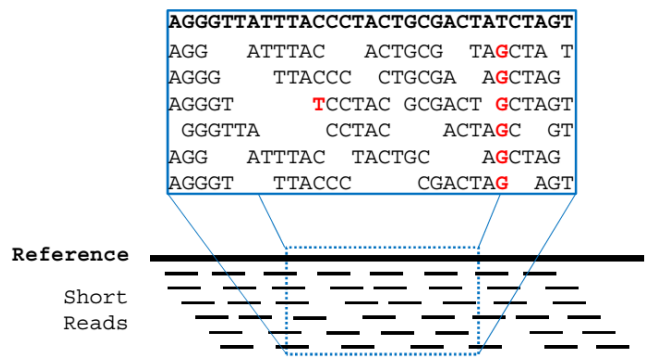

Fig. 1. Short reads aligned to the reference genome. Differences are shown in red, including a SNP (G for T) and a read error (T for C).

## 2.2 Short Read Alignment

Among the computational steps needed to make sense of NGS data, one of the most significant is alignment. NGS machines are often used for genome sequencing. In these cases, the user typically wants to know something about how the sequenced individual differs from a known reference genome. NGS introduces significant computational challenges to the relatively simple problem of genome alignment. First, as shown in Fig. 1, the short reads may contain both errors in an individual read (introduced by the sequencer, the replication process, or another source) and actual differences between the individual sequenced and the reference genome. Additionally, because the short DNA fragments are not sequenced in any particular order, the short reads arrive from the sequencer without any information about the part of the genome to which they align. Potential errors and differences from the reference cannot be determined without first knowing this alignment, so a process called short read alignment is required.

Through alignment, each short read is assigned zero, one or more potential alignments, as well as a score attempting to quantify how likely that alignment is to be correct. Given the errors introduced in the sequencing processes and the extremely repetitive nature of some portions of the genome, the accuracy of these alignments varies significantly. Short read alignment is typically one of the first steps of a long genomics pipeline, which must potentially process a few billion reads for one individual, so quickly achieving accurate alignments with good information about when an alignment is unreliable is critical to the performance of the pipeline. Because the cases where an individual's genome differs from the reference are often the significant ones, the alignment algorithm must be able to tolerate some difference between the short read and the reference. Later pipeline stages typically evaluate all of the reads aligned to a particular reference base to determine which reads contain errors and which describe actual genetic variation.

## 2.3 BFAST

There are a number of software packages that perform short read alignment. These include BFAST [5], BWA-SW [6], Bowtie2 [7], GSNAP [8] and SOAP2 [9]. These tools represent a number of tradeoffs between accuracy and performance [10]. The initial version of our FPGA-based aligner, described in the next section, uses some techniques from the BFAST aligner, described in this section, although it employs a different scoring algorithm.

The fundamental problem faced by short read aligners is that the errors discussed above typically occur at a frequency of one, two or more per read, with longer reads tending to have more errors per nucleotide. For this reason, as well as the existence of individual variation from the reference, most commonly in the form of Single Nucleotide Polymorphisms (SNPs), exact string matching cannot be used to align short reads to a reference. Algorithms that find alignments while tolerating errors, such as Smith-Waterman, are much slower, and thus it is typically not practical to compare each short read to the entire reference genome (over 3 billion nucleotides for humans).

To address this problem, we adopt a strategy pioneered by BFAST and other aligners. Alignment is performed in two stages. The first winnowing stage identifies promising candidate loci in the reference sequence. This is done by splitting the short read into smaller sequences of length k and identifying reference locations that are exact matches to these subsequences, or seeds. This is done quickly using a fast exact-matching index lookup, resulting in one or more Candidate Alignment Locations (CALs) in the refer-ence for each read. The index is precomputed from the reference and does not have to be regenerated for each run. This approach works because as long as the seeds are short enough relative to the read, some of them should be error- and variation-free and thus match the reference exactly.

The second stage scores the alignment of the entire read against each CAL using a slower but more powerful method that can tolerate some errors and variation while still identifying likely matches (Smith-Waterman in our work). Because of the high level of filtering provided by the first stage, this can be done in a reasonable amount of time. For example, with reads of length 100 and seeds of length 20 there will be some seeds that are exact matches to the reference, despite a few differences per read due to errors or individual variation. Most of these length 20 seeds are sufficiently unique that there will be a limited number of CALs that require exact alignment for this read. This introduces a number of obvious tradeoffs in the areas of speed, memory and alignment accuracy. Smaller seed lengths will increase sensitivity and error tolerance, but also increase the number of CALs for each seed since shorter strings repeat more often in the genome. Similarly, using overlapping seeds improves error tolerance but also significantly increases the number of seeds that must be examined for each read (in our example there are 5 non-overlapping seeds per read but 81 overlapping seeds.) There are also significant engineering tradeoffs to be made in compactly storing and efficiently accessing the index data structure. Much of the rest of this paper is devoted to exploring these tradeoffs and explaining why they were made as they were with a sensitive and flexible FPGA accelerator in mind. We will begin with a brief discussion of how BFAST approached these issues.

The BFAST index was designed to use large search keys (seeds) to attempt to find as close to a unique genome location as possible (one CAL per read) while still allowing for an O(1) lookup for each key [5]. In order to achieve this, BFAST creates a number of indexes from the reference genome, each using a different (generally not contiguous) subset mask. These masks provide gaps that can ignore different SNPs and read errors. Because CALs are only generated from a portion of the read that is an exact match with a masked portion of the genome, using a larger key size results in fewer CALs per read but can also reduce the sensitivity of the matching in the case of read errors or SNPs.

In order to get only a few CALs per index per read, BFAST actually uses k = 22 for the human genome (k is the k-mer length, i.e. the length of the substrings selected from the read). Note that for large k's like this, all possible k-mers do not occur in the genome. For this reason, an index of all k-mers would be very sparsely populated and inefficient. Instead, BFAST uses indexes of only k-mers that actually occur in the genome and accesses them using a dense table of the first j bases of the k-mer being looked up, where j < k. For example (see Fig. 2), for the sequence "ATGACGCA" and the mask 101101 with k = 4 and j = 2, the 2-mer "AG" would be looked up in this dense table. For the human genome and a k of 22, BFAST uses j = 16 so that this table remains manageable at 4G entries. Each entry in the table provides a position in the index for the first and last position of k-mers that start with the j-mer, and because the k-mers are sorted in the index, a binary search can very quickly find the desired k-mer and therefore the relevant CALs. The two stage CAL lookup process described above provides one inspiration for our system but lacks flexibility as discussed below.

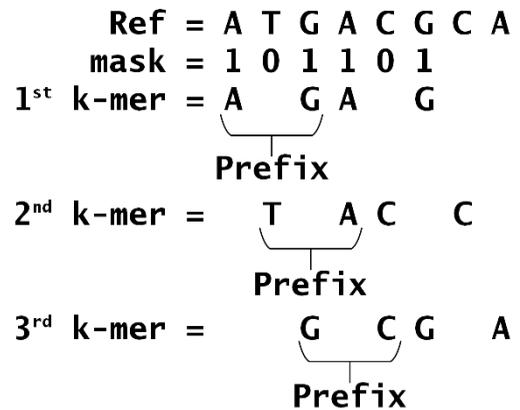

Fig. 2. Simplified BFAST index construction example, where k = 4 (key size) and j = 2 (prefix size).

## 3 SUMMARY OF PREVIOUS FPGA DESIGN

This section will summarize the architecture of the first version (V1) of our hardware accelerated short read aligner, described in detail in [11]. The V1 system, inspired by BFAST, implemented a short read alignment pipeline primarily on FPGAs. This pipeline accepts a stream of short reads and outputs these reads aligned to the reference genome using the following modules: Index Construction, CAL Finder, CAL Filter, Reference Retrieval, Smith-Waterman Aligner, and Score Tracker.

### 3.1 Index Construction

Index construction is not an FPGA task, but rather a step that must be done before running the FPGA algorithm. This step is performed on a host machine and need only be run once for each reference and seed parameter combination. Thus, the index would generally be pre-computed when a user's computation methodology was defined, and used from then on, potentially for weeks or months. The index consists of two tables: the pointer table and the CAL table. These tables together indicate, for a given seed, the location of all occurrences of that seed in the reference genome. Thus, for a seed length of 8, there would be an entry for all occurrences of "AAAAAAAA", another for "AAAAAAAC", etc.

In order to construct these tables, the entire reference genome is broken up into overlapping seeds. See Fig. 3 for an example of this process. In the case of the V1 system, the seed length was configurable, but typically set to 22 bases. Each seed is further subdivided into address and key section, as shown in the figure. The address is used as the index into the pointer table; the length of the address determines the size of the pointer table because the pointer table has one entry for every possible address. Since there are four bases, this works out to $4^{\text{address\_length}}$ entries. Note that there will be address table entries for every possible string of the address length, even those that do not show up in the reference. For this reason, the pointer table is a fixed size for a given seed and key length.

```
Ref   = T A C A G C A T A C G G T T T T G T T C T C G C
Seed0 = T A C A G C A T A C G G T T T T G T T C T C
Seed1 =   A C A G C A T A C G G T T T T G T T C T C G
Seed2 =     C A G C A T A C G G T T T T G T T C T C G C
      =     0100100100110001101011111111011110111011001
```

Fig. 3. The seeds for a reference sequence, with seed 2 divided into address bits and key bits.

The V1 system uses a 22 base seed for the same reasons as BFAST. This length provides good specificity, avoiding too many false positives in the form of numerous extraneous CALs for each read, while also being short enough to allow many mismatches per read. At 22 bases for non-overlapping seed, an 88 base read with fewer than 4 errors is guarteeed to have at least one perfectly matching seed and is therefor likely to identify some CALs. However, as with BFAST, the data structure implied by a 22 base seed would be very large ($2^{44} = 1.76 \times 10^{13}$ entries) and mostly empty, since many seeds never occur in a particular reference genome or, for that matter, in the genome of any living organism.

To solve this problem, the V1 system uses two tables. The CAL table contains <key, CAL(s)> pairs. Each of these entries stores all of the one or more CALs associated with a seed actually found in the reference. If the key in the CAL table matches the key bases from the end of the seed (7 bases in this example), that CAL is considered a match. This means the relevant portion of the reference should then be looked up and scored, as discussed later.

Although the CAL table contains CALs for each sequence in the reference, it will not contain entries for all possible sequences. Instead, a second table, called the pointer table, is used to index into the CAL table. The pointer table contains one entry for each possible address where an address is a fixed length prefix of the seed (15 bases in the example from Fig. 3). Lookups into the pointer table using the address are fast, and the pointer table contains pointers into the CAL table or a null entry if the reference contains no sequences matching the address. The V1 system provides additional features to compress the table and improve lookup speed, but these are beyond the scope of this paper.

To illustrate this (see Fig. 4), with a shorter seed to improve clarity, if the seed were "TACACGTA" the first entry in the address table would be accessed. The base pointer points to position 709 in the CAL table. Finally, the key at position 711 in the CAL table matches the suffix of the seed "CGTA", so the seed matches position 5037 in the reference. Note that multiple CALs for the same key are possible, such as the two CALs for seed "TACACATA".

Pointer Table

| Addr | Base Ptr |
|------|----------|
| TAA  | 709      |
| TAC  | 712      |
| TAG  | 713      |
| TAT  | null     |

CAL Table

| Addr | Key | CAL |
|------|-----|-----|
| 709  | CAG | 4977 |
| 710  | CAT | 3520 5642 |
| 711  | GGT | 5037 |
| 712  | AGC | 6179 |
| 713  | ACT | 1635 5484 7212 |
| 714  | CCT | 2823 |

Fig. 4. An example of pointer and CAL tables.

An additional complexity is introduced to this table by the fact that any given read may be from either the forward or reverse strand of the DNA double helix. Because the reverse strand is made up of the complementary bases of the forward strand, but running in the opposite direction, we call this the reverse complement. For example, if the read is "ACC", the reverse complement is "GGT". Since there is no way to know which strand the short read is from, the system must be able to handle either seamlessly. This is done by generating the reverse complement of each seed and only looking up the lexicographically smaller of the two. This operation works correctly regardless of the original strand, since the reverse complementing operation is symmetrical. A similar operation is used when generating the tables, so only CALs from the lexicographically smaller seeds are present. The only additional complexity then required is keeping a bit with each CAL to signal if it is from the forward or reverse reference and performing the final Smith-Waterman operation against the reference in the correct direction.

In the basic organization described above, the number of CAL table entries associated with each pointer table entry would vary wildly due to the non-uniform distribution of seeds in human or other DNA. To avoid this problem and provide a more even distribution of CAL table buckets, the entire seed is hashed. With this hash in mind, the entire process consists of comparing the seed to its reverse complement, hashing the smaller of the two seeds, looking up the correct pointer table entry using the address bits, following the pointer table to the correct CAL table bucket, and then checking each key found there against the key from the seed. This last step requires walking the entire region of the CAL table specified in the pointer table to check for CALs with matching keys.

### 3.2 CAL Finder

The short read pipeline relies on a preconstructed index, built from the entire reference genome or some other sequence, as described above. Each read is divided into many overlapping or non-overlapping seeds depending on the biologist's preference (this is different than index construction, where overlapping seeds are always used). In the case of overlapping seeds, every possible full length seed is generated from the read for a total of (read length – seed length + 1) seeds. In the case of non-overlapping seeds, every full length seed is generated, starting at the beginning of the read, such that no base is in multiple seeds. This results in ⌊read length / seed length⌋ seeds, and if the read length is not divisible by the seed length there will be some bases at the end of the read that are not part of any seed. An intermediate approach, with seeds overlapping by fewer bases, is also possible.

Each seed is looked up separately using the pointer and CAL tables. For the V1 system, overlapping seeds of 22 bases are used, and the CAL Finder module must walk the pointer table and CAL table as described above. The CAL Finder has one FPGA DRAM port for pointer table access and another for CAL table access. After loading the correct portion of the CAL table from DRAM, the CAL Finder sends all of the CALs with matching keys to the next module in the system (see Fig. 5).

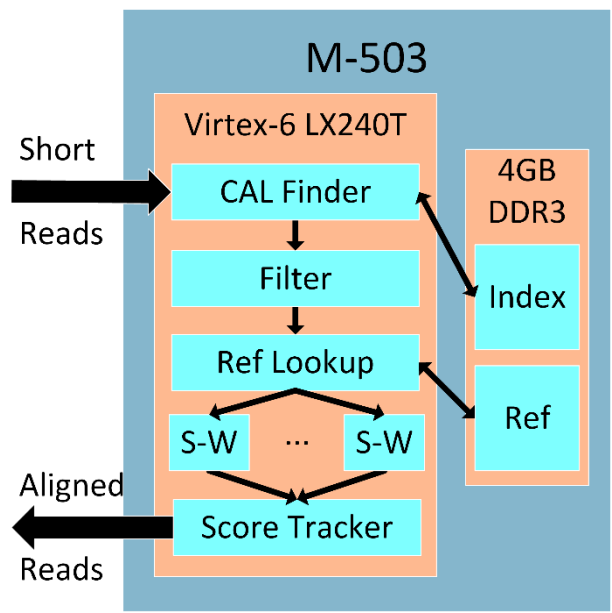

Fig. 5. Architecture of V1 short read alignment system.

### 3.3 CAL Filter

The CAL Filter has two functions. First, the CAL Filter combines duplicate CALs into a single reference lookup to avoid wasting DRAM bandwidth and S-W resources. This is necessary because usually many of the CALs for seeds of a given read will be near each other in the reference. For an exact match with overlapping seeds, each seed will return a CAL one base after the previous seed. Second, the CAL filter translates the CAL to determine which 256-bit DRAM words of the reference must be read. Depending on the length of the short reads (typically 76 bases for V1) and the way the CAL falls relative to the start of a DRAM word, one or more words may be required for each CAL. After the CAL filter determines which reference words are required, the reference retrieval submodule loads them from DRAM, using a single port. For reads that have CALs that are not near each other, this may require loading many non-contiguous portions of the reference.

### 3.4 Reference Retrieval

This very simple module manages memory accesses on the FPGA, to keep the alignment units supplied with the appropriate sections of reference genome as specified by the CALs.

### 3.5 Smith-Waterman Aligner

Once the reference is retrieved, each read-CAL pair is passed to a Smith-Waterman (S-W) aligner, along with the appropriate reference. Unlike the other modules, which exist in a single chain in the V1 system, there are multiple S-W aligners in parallel to help alleviate what would otherwise be the performance bottleneck. The aligners in V1 use Smith-Waterman with the affine gap model, which allows the short read to have long insertions or deletions of bases compared to the reference without a significant score penalty. This scoring allows for more biologically relevant results. Details of the dynamic programming algorithms used in scoring can be found in [12].

The S-W implementation uses a 1-D systolic array of processors, which parallel-loads each short read (one base per processor) and matches it against a streaming reference as shown in Fig. 6. The S-W algorithm requires a 2-D dynamic programing table, where the vertical axis represents the sequence of the read and the horizontal axis represents the sequence from the reference. The algorithm can calculate the value of a cell in the table using only the adjacent cells above, to the left, and the up-left diagonal cell. This means that the 1-D array of processors in each S-W unit can parallel load the read and then stream the reference sequence through them, producing a computation wavefront along the antidiagonal of the table, progressing from left to right. The cell with the largest score in the final row of the table represents the best score for the entire read, and is the output of the aligner.

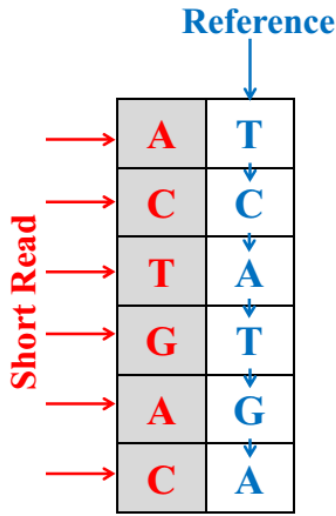

Fig. 6. Smith-Waterman dynamic programming table (top) and hardware array based implementation (bottom).

### 3.6 Score Tracker

After an S-W score is computed for each unique CAL, the Score Tracker picks the reference position that has the best score for the read and is therefore the closest match. The best score may not start precisely at the candidate location, but may instead be off by a number of bases bounded by the read length (in the case where the seed is close to the end of the read). This is determined by the highest scoring cell in the final row of the S-W dynamic programming table as discussed above. There are multiple modes of operation available in V1, but typically only the best scoring result for each read is returned from the FPGA to the host, while the others are ignored.

### 3.7 V1 FPGA Implementation

The V1 implementation uses Virtex-6 LX240T FPGAs on Pico Computing M-503 modules [13], discussed in more detail later. Each module is equipped with two 4GB DDR3 DRAMs running at 400 MHz. This DRAM is used to store the index and reference. For the human genome, the memory requirements are as follows:

$$\text{reference} = \sim 3.2 \text{ Gbases} * 0.25 \text{ bytes/base} = 0.8 \text{ GB}$$

$$\text{pointer table}$$
$$= 2^{\text{address bits}} \big( \text{start pointer bits} + 2^{\text{tag bits}} * \text{offset bits} \big) =$$
$$2^{26}(32 + 2^4 * 14) = 2 \text{ GB}$$

$$\text{CAL table} = 1 \text{ CAL/base} * \sim 2.44 \text{ Gbases}$$
$$* (4\text{-byte location} + 4\text{-byte key})/\text{CAL}$$
$$= 19.5 \text{ GB}$$

This results in a total memory requirement of over 23 GB, when using 26 address bits or 13 address bases (the default value for V1). The CAL table has fewer Gbases than the entire reference because of unkown bases or Ns.

As these calculations show, the size of the CAL table is much larger than the DRAM available on an M-503, so for V1 the design was partitioned across multiple modules. This means that each module was given a subset of the pointer table and CAL table, split up using the first n bits of the seed. For example, using the first 3 bits of the seed, the index could be partitioned across 8 FPGAs, with each FPGA receiving 0.25 GB of the pointer table and 2.4 GB of the CAL table. This, along with the complete reference (which is required on each card) results in a total memory usage of only 3.45 GB, while also improving effective memory bandwidth across the system. When a read is streamed to each of the eight FPGA modules, the CALs for that read can be looked up in parallel across all of the modules. As discussed in the next section, memory bandwidth is a significant bottleneck for the V1 system, so this partitioning is desirable.

The CAL Filter is implemented using a hash table made up of FPGA block RAMs as described in [11]. Each FPGA is given as many S-W modules as can fit to process all of the CALs being returned from the CAL Finder. All of the modules run at a 250 MHz clock, except the S-W processors which can only run at 125 MHz due to the significant sequential computation required.

### 3.8 Performance Considerations

In the V1 system, the S-W modules are fully pipelined during computation, but each module can only work on a single short read and reference segment at a given time. For efficiency and simplicity, the V1 system always processes the reference in 256-bit DRAM word sized sections. When the relevant reference from the CAL falls across a DRAM word boundary the S-W module must align the read against 512-bits of reference, or 256 bases. Because each read is loaded in parallel the engine is not pipelined between different reads, so there is a lead-in time equal to the length of the read, in this case 76 bases. The total time to process one reference section for one read is 256 + 76 = 332 cycles at 125 MHz or 377 kCALs / sec per S-W module. In the best case, assuming many CALs per read, performance could approach 256 cycles per CAL or 488 kCALs / sec per S-W module.

The V1 memory controller on the FPGA is clocked at 200 MHz and requires 45 controller clocks for each access, resulting in about 4.4 M transfers / sec. In this system, the memory accesses are not pipelined. Because of this, for each read the CAL Finder requires the full time to access the pointer table for each seed, followed by accessing the CAL table for each seed. Assuming an average of about 8 CALs being found for each read, as in [11], that also requires 8 reference loads per read. Accessing the pointer and CAL tables once per seed results in (76 base read - 21 incomplete seed positions for a 22 base seed) * 2 + 8 reference access = 118 total memory operations per read. At 4.4 Mops this results in 37 kreads / sec, or 75 kreads / sec given the two independent memory channels on the M-503. However, given the partitioning scheme described above, a four M-503 system like V1 would only have to look up ¼ of the seeds for each read on each FPGA, effectively quadrupling performance while also allowing for the entire human genome and associated tables to fit in M-503 DRAM. The total performance of this system would be memory limited at 300 kreads / sec.

With this rate of memory access, only two S-W engines are required per FPGA to fully saturate the memory. Given that the V1 systems can fit a maximum of 6 S-W modules per FPGA, it is clear that the system is memory bandwidth limited. Pipelining the memory access portion of a 4 M-503 V1 system to saturate 6 S-W modules per FPGA would require 380 kCALs / sec / S-W module * 1 read / 8 CALs * 6 S-W modules / FPGA = 285 kreads / sec / FPGA. Given the current memory limit of 75 kreads / sec / FPGA, it would take memory pipelining with 4 reads in flight at a time to fully saturate the memory system on an FPGA with 6 S-W modules. Although this option is not available in the V1 system, such a memory controller is feasible and could result in a system bottlenecked by FPGA resources instead of memory bandwidth.

## 4 DESIGN CHANGES

To summarize the previous section, the bottleneck of the V1 system is the memory operation rate on the FPGA boards. This bottleneck could be alleviated by pipelining the CAL Finder's DRAM interface, allowing for better utilization of the available memory bandwidth. This change would result in the CAL Finder and Filter modules taking up more FPGA resources, and those limited FPGA resources would become the bottleneck by reducing the number of S-W modules that could fit on the FPGA. Additionally, no amount of pipelining can overcome the fact that the aggregate memory bandwidth on the FPGA board is significantly less than the multiprocessor host system. These statements are analyzed in more detail in Section 6 of [11] and will be discussed further below.

Even disregarding the performance, the V1 system has additional disadvantages. Because the CAL Finder and CAL Filter are FPGA modules hand coded in Verilog, any changes to this logic tend to be time consuming and error prone. Furthermore, although index construction itself is performed on the host, the CAL Filter is hardcoded with many of BFAST's assumptions related to the seed and key length, as well as the maximum number of CALs per seed and total genome size. These restrictions are particularly problematic for biologists who are accustomed to making frequent small changes to their algorithms. The appeal of an accelerated solution is significantly lessened if it is brittle and difficult to modify. Additionally, biology is full of problems that require finely tuned sensitivity parameters to achieve acceptable quality of results.

In order to address all of these issues, the current version of the system (V2) was designed to use the host for index lookups before sending the reads and all matching CALs to the FPGA for alignment. The design of this system can be seen in Fig. 7.

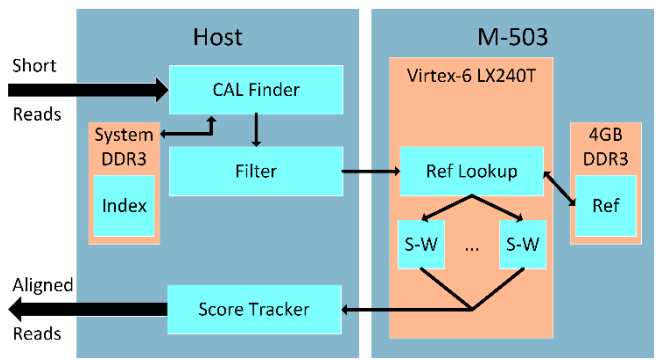

Fig. 7. Architecture of V2 short read alignment system.

## 4.1 Index Construction

Index construction is similar to the V1 system. However, because the index lookups are performed in software, most index features can be configured at compile time or runtime. In addition to the ability to configure the number of bases that make up the seed, the breakdown of the seed into address and key bits is completely configurable. Increasing the seed length reduces runtime, since fewer seeds are generated from each read, but also reduces sensitivity. Increasing the portion of the seed dedicated to address bits results in a larger pointer table, since the pointer table is directly indexed. The advantage of this is a reduction in the time required to scan the CAL table, since there will be fewer non-matching CALs in each pointer table slot. All seed bits that aren't used for address are used for the key, to match CALs in the CAL table. Changes to these values allow the system to mimic the results of other software packages more closely. For example, although they are similar in some ways, mrFAST and mrsFAST use very different seed configurations than BFAST [14], [15]. Being able to easily support both is a significant advantage over the V1 system.

In addition to the seed, the all-software indexing solution allows for configuration of the size of fields in the tables. Because the tables are loaded from disk for each alignment, different table configurations can be used with indexes built from different reference genomes without recompiling software or reconfiguring the FPGAs. This was not possible with the V1 system. This configuration includes the number of bits at each offset position. Increasing this value increases the size of the pointer table, but allows for more CALs at each offset position. As in the V1 system, the maximum number of CALs for a single seed is configurable, but it can now be set on an index by index basis.

## 4.2 Other Modules

The V2 CAL Finder and CAL Filter modules are implemented as software threads that communicate through Liberty Queues [16]. The organization of threads is shown in Fig. 8. The greater aggregate memory bandwidth available on the host machine (section 5.1) to perform index lookups depends on an intelligent organization of threads to hide memory latency and achieve the maximum memory operation rate. As discussed below, the thread structure used here is sufficient to provide CALs to up to 4 FPGAs without becoming the bottleneck. A system that used C-slowing or some other technique to fit more S-W engines on the FPGAs could require more memory bandwidth on the host to keep up.

To summarize, a single thread reads the short reads from disk, and dispatches each read to a pool of index threads in a round robin fashion. Each index thread pair is assigned to an individual FPGA, and both index threads write to a single FPGA control thread that performs the same operations as the CAL Filter described in the previous section. These threads are paired to optimize memory access for our system, in which memory is local to one of two CPU sockets, and could be configured differently in a different system. The index thread sends the short read, followed by the filtered CALs for that read, to the FPGA; a similar thread receives results from the FPGA. The receive thread performs the operations of the Score Tracker, but is significantly more configurable than the V1 firmware version. The filtering options include returning the single highest scoring CAL, some number of highest scoring CALs, or all CALs tied for the highest score. This thread also formats the output into a standard format like SAM [17]. This output is then written to one file per FPGA, to be aggregated later if required.

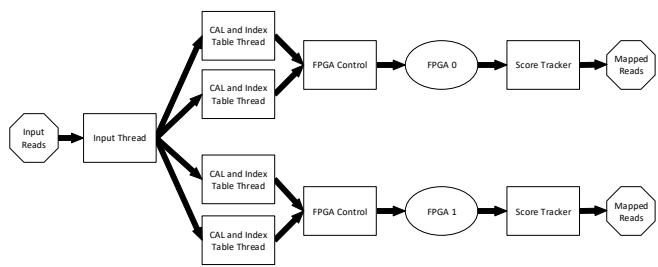

Fig. 8. V2 software thread setup for 2 FPGAs.

## 4.3 V2 FPGA Implementation

The V2 implementation uses the same FPGA boards as V1. Because only the reference has to be stored in DRAM on the FPGA modules, no partitioning of the reads is required for a human sized genome. CALs from any read can be aligned on any FPGA in the system. Software and HDL source code from the V2 aligner is available at http://students.washington.edu/nmcvicar/aligner/.

## 5 PERFORMANCE COMPARISON

Besides flexibility, the primary advantage of V2, compared to V1, is performance. As described above, the performance of the V1 system is limited by FPGA to DRAM bandwidth. Moving the CAL and pointer tables to the host, where DRAM bandwidth is greater, alleviates this bottleneck. This, in turn, frees up FPGA resources to create additional S-W units on the FPGA, should that become the factor limiting performance. The baseline V1 system can process between 150k and 200k full human genome reads per second per FPGA, as described in more detail in [18]. The V2 system can do significantly more than this, as described below.

## 5.1 FPGA Platform

In addition to performance, the V1 system's use of FPGA-attached DRAM for reference and index storage introduces another challenge. As discussed previously, the V1 system is implemented using Pico Computing M-503 FPGA modules as well as M-501 modules [13]. The M-501 implementation is described in detail in [11]. Significantly, the M-503 module has two 4 GB DDR3 DIMMs while the M-501 has only 512 MB of DRAM. The human genome requires CAL and pointer tables totaling 21.5 GB, in addition to the reference (3.2 Gbp or 800 MB). This means that depending on the partitioning, 48 M-501s might be able to align reads to the full human genome, but fewer would not. 48 is the maximum number of M-501 modules a 4U Pico Computing system can support using 8 PCIe slots with 6 M-501s on each EX-500 backplane.

For the current system, the tables are stored on the host. However, the 512 MB on the M-501 would still limit how much of the human genome reference could be stored near each FPGA. Furthermore, some applications might require larger references, including the case study discussed below. Partially to address this issue, the V2 system is implemented exclusively on Pico Computing M-503 modules. These include the same Virtex-6 LX240T FPGA, but have 8 GB of DRAM per FPGA. This allows the system to store the complete reference genome with each FPGA, even for larger references. The theoretical maximum bandwidth of the M-503 DDR3 is ~10.8 GB/s, compared to a theoretical maximum of ~34 GB/s and measured performance of over 27 GB/s on our host.

## 5.2 Performance Improvements

For testing purposes, our system contains two Nehalem E5520 processors (8 cores total), 48 GB of DDR3-1333 DRAM and two Pico Computing M-503 modules. Although the specifications of this system are not identical to those used in [18], this has no impact on the evaluation because the V1 system is almost entirely reliant on the FPGAs for performance. The FPGAs are the same, as is the 50 million short read sample dataset.

|  | V1 System | V2 System |
| --- | --- | --- |
| CALs / s / FPGA | 632 k | 2231 k |
| System Power (W) | 435 | 384 |
| Energy / CAL (J) | 688 μ | 172 μ |

In addition to the improved flexibility, the V2 system is significantly faster than V1 (see Table 1). Note that this table measures performance in CALs / sec and not reads / sec as used above, since users can adjust CALs / read to tradeoff between sensitivity and performance.

The performance improves by a quantity very similar to what was estimated for the "CALFinder on Host" system described in [11]. This analysis was performed by estimating the number of seeds the host could generate per second based on the number of clock cycles required to generate a seed and the number of memory references per second supported by the host. If the host can generate enough seeds, the total CAL alignments per second supported by the FPGAs gives the total for the system. Because our current system only has two FPGA modules, these results are computed per FPGA in order to allow a direct comparison with [18]. The V2 system is 3.53x faster and 4.00x more energy efficient than the V1 system. The exact improvement is highly dependent on the exact reference size and portioning used, as well as the number of CALs per read, but substantial improvement is seen across the board. One of many uses for the ability to carefully tune CALs per read will be illustrated by the following case study.

Understanding the CALs / secs performance metric requires a brief discussion of the variability and distribution of the number of CALs / read. In particular, this number is highly dependent on both the seed and key length as well as the source of the reads. The frequency and length of repetitive regions is highly variable across genomes and within a single genome, as illustrated in Fig. 9.

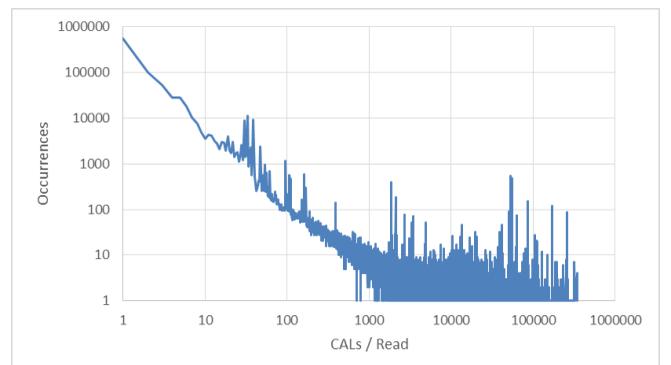

Fig. 9. CALs / read distribution for 1 M human RNA-Seq reads aligned to HG19.

On the left side of the graph the vast majority of reads only result in one or a few CALs. This makes intuitive sense because neighboring CALs are combined so a read that maps to a single unique region of the genome will only send one CAL to the FPGA for scoring. As the number of CALs / read increases the frequency decreases exponentially, but there is a very long tail where the number of CALs / read is extremely high and these occur more often than would be expected if one were to only look at the left half of the graph. These are the highly repetitive regions of the genome where a small string of bases is repeated many times. Some aligners limit the maximum number of CALs for a single seed, reasoning that for reads that map to so many places in the reference it is highly unlikely that one mapping will be significantly better than any other. Our system supports indexes built in this manner, and in fact gets a significant performance boost from doing so, but in the interest of doing an equal amount of computation we have not enabled this for the above comparisons to the V1 system. Given the fact that our current system is CAL scoring limited when using a single FPGA, it is clear that much of the variation in reads / sec measurements is due to changes in CALs / read. In other words, by limiting CALs / seed the number of CALs the system must process for each read decreases significantly (by more than an order of magnitude) and performance measured in reads / sec increases by roughly the same factor, as seen in the case study below.

To summarize, moving the CALFinder and CALFilter from the FPGAs to the host system, freeing significant FPGA resources for S-W alignment and removing all traffic from the FPGA DRAM except for reference lookups, offers performance improvements as well as significant benefits to configurability and genome support.

## 6 CASE STUDY

Our development of the V2 system was driven by two goals: (1) improve overall performance by moving most memory accesses to the higher-bandwidth host memory; (2) improve flexibility of the system to adapt to the wide range of usage patterns needed by working biologists. In the previous sections we have discussed the performance improvements achieved in V2. In this section we present a case study featuring a use of short read alignment that is important to biologists and possible in V2, but would have been too complicated with too large of a reference genome to consider performing with V1.

### 6.1 Mouse Human Coalignment Motivation and Challenges

So far in this paper we have discussed DNA sequencing, the process of identifying the sequence of bases found in the genome of a given organism. A related technique is RNA-seq, the process of identifying the sequence of bases found in the RNAs currently active in that organism. RNAs are short strands of DNA-like bases that perform important functions in the cell, and the abundance levels of specific RNA sequences are biologically important. Like DNA, RNAs are sequences of 4 possible bases, and are derived by copying segments of the organism's DNA called genes. However, the RNA molecules are quickly edited ("spliced") to remove extraneous segments ("introns") that separate functionally important segments ("exons"). Furthermore, alternative splicing causes some exons to be optionally retained or deleted (see Fig. 10). The different alternative splicings of a single gene are called isoforms. Most genes are spliced and most spliced genes exhibit multiple isoforms [19].

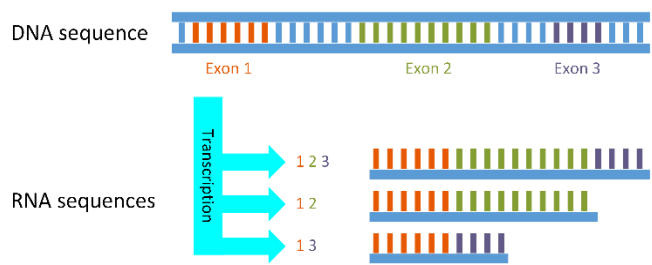

Fig. 10. A DNA sequence with three exons and three possible alternatively spliced RNA products.

RNA-seq can be performed via similar processes to DNA sequencing, resulting in a large number of short reads of RNA data that need to be aligned to a reference genome. One could perform this alignment to the source DNA sequence, but the introns and alternative splicings will result in very poor matches for many sequences. While de novo isoform discovery from RNA-seq data is desirable, it is computationally expensive and potentially inaccurate, so a common alternative is to construct a reference that directly includes all known isoforms. Although this will significantly increase the size of the reference, it can also provide significantly better alignments.

RNA-seq can be used for many different applications. One such use is to determine the source organism when RNA from multiple different species may be present in a sample. For example, in tumor xenografts, human tumor cells are grown in a mouse host, enabling detailed studies of human tumors in living hosts that would not otherwise be possible [20]. However, RNA extracted from the xenografts will contain RNA from human tumor cells as well as contaminating mouse RNA. Thus, the RNA-seq data that is obtained needs to be sorted into relevant human RNA and contaminating mouse RNA. A conceptually simple approach for attempting this is to align each short read to both the human and mouse references, and to classify each as human vs mouse based on which reference yields the higher alignment score.

## 6.2 Coalignment Experiment Design

To test our aligner against existing approaches, we will apply it to categorizing human and mouse RNA in an RNA-seq application: analysis of human retinal stem cell derived tissues following transplant to a mouse or culture of mouse cells [21][22]. This problem is challenging because the RNA-Seq data, which must be collected from more than a single cell, will include reads for both mouse and human genes. To provide a dataset where we know what the "right" sorting is, we simulate this scenario using a set of human reads from fetal human retinas and a set of mouse reads from cultured mouse retinal cells.

Human fetal tissue was obtained from the Birth Defects Research laboratory at the University of Washington without identifiers. Mice were housed in the Department of Comparative Medicine at the University of Washington. All procedures were carried out in accordance with protocols approved by the UW IACUC. Total RNA was isolated from postnatal day 0 mouse retinas and 80d whole human fetal retinas using Trizol (Invitrogen, GrandIsland, NY). RNA was treated with RQ1 DNase (Promega, Madison, WI) for an hour and then purified using miRNeasy kit (Qiagen, Germantown, MD). 5 µg of each sample was then processed using Ribozero (ScriptSeq Complete kit; Epicenter, Illumina) to deplete ribosomal RNA. Libraries were generated and sequenced on an Illumina HiSeq to produce 100 base pair long paired end reads.

Each set of reads was aligned to both the mouse and human genome, with and without alternative splicings included with the reference. The alternative splicings used were from Ensembl release 72 and increased the total human reference genome size by 9% by adding 195k individual isoforms. Aligning to both genomes allows us to evaluate and compare the accuracy of the alignments, since it is reasonable to expect that the human reads will align better to the human reference and the mouse reads to the mouse reference. Analysis validated this assumption, so for the following evaluation we predict a read to be human if it has a higher score against the human reference genome and mouse if it has a higher score against the mouse reference. Furthermore, for these experiments we report results only for reads from human retina cells aligned to both the human and mouse reference. The results for the mouse reads are similar enough to be considered redundant within the scope of this paper.

## 6.3 Software Aligners

To evaluate our system, we compare its performance to three software based aligners. These are BFAST [5], GSNAP [6] and BWA-SW [6]. BFAST has already been discussed in detail. GSNAP is an aligner specifically designed to get good results in the case of alternative splicing events, as well as supporting common biological variation better than BFAST. Finally, BWA is a Burrows-Wheeler transform based aligner that has traditionally been extremely fast, although it has a poor quality of results in the face of variation or large read errors. More recently, an enhanced version, called BWA-SW, has been produced which uses a Smith-Waterman aligner in the final stage, significantly improving variation tolerance. BWA-SW is not specifically designed for alternative splicing, but like our FPGA system it can handle very large references without trouble, so the same method of including each isoform can be used.

## 6.4 Results

The results of the first experimental runs, without alternative splicings, are shown in Table 2. Each row contains the results for an aligner, and the columns show the percentage of reads that were either correctly identified as being of human origin or incorrectly identified as having a mouse origin. A correct identification comes from aligning the read to both the human and mouse genomes and having the alignment to the human reference scoring better than the alignment to the mouse reference. The last two columns show ties, where the read aligned equally well to both references and unaligned reads which could not be aligned to either reference with any reasonable level of certainty. The case where a read aligned to one reference but not the other was marked as correct or incorrect.

TABLE 2
ALIGNER QUALITY RESULTS (HUMAN READS)
WITHOUT ALTERNATIVE SPLICINGS

| Aligner | Correct | Incorrect | Tie | Unaligned |
|---------|---------|-----------|-------|-----------|
| BFAST | 86.13% | 8.00% | 5.87% | 0.00% |
| GSNAP | 75.2% | 0.03% | 18.2% | 6.36% |
| BWASW | 93.8% | 1.23% | 3.14% | 1.83% |
| FPGA | 93.3% | 2.57% | 3.32% | 0.80% |

When alternative splicings are not included, the results are still very good overall. BFAST, which is the oldest aligner, unsurprisingly gets the worst results in terms of incorrect classifications. It is included here because it uses techniques somewhat similar to our FPGA aligner, as described above. GSNAP has many more ties than the other aligners, presumably due to its coarser grained scoring system. The results are very similar between BWA-SW and the FPGA aligner. Both use S-W for scoring and were configured to use the same S-W mismatch penalties, so this is an expected result. Without alternative splicing, BWA-SW has fewer incorrect alignments but more reads going unaligned. This is likely due to the FPGA system's more permissive CAL Finder, which can handle more mismatches than the BWT based initial matching in BWA-SW.

Table 3 shows data for the case where alternative splicings are included in the reference. The FPGA-based system has the most correct and fewest unaligned results, although GSNAP has the fewest incorrect alignments while still producing many ties. BFAST, which is very sensitive to reference size, cannot handle a reference as large as is needed to include the alternative splicings with default settings. Overall, the FPGA system correctly classified 1% more of the reads than BWA-SW while also classifying 0.29% fewer incorrectly and aligning 1.05% of reads that BWA-SW was unable to. Although the FPGA system had slightly more ties, a tie is a much better result than an unaligned read because a read that didn't align to either genome provides no information to downstream processing modules. A tie aligned to both genomes with the same score, and can often be resolved later using the paired-end read or other methods. Taking all of these results in conjunction, the FPGA system produces results of noticeably better quality than BWA-SW when alternative splicings are included.

TABLE 3
ALIGNER QUALITY RESULTS (HUMAN READS)
WITH ALTERNATIVE SPLICINGS

| Aligner | Correct | Incorrect | Tie | Unaligned |
|---------|---------|-----------|-------|-----------|
| GSNAP | 77.8% | 0.49% | 20.6% | 1.04% |
| BWASW | 94.4% | 1.01% | 3.04% | 1.53% |
| FPGA | 95.4% | 0.72% | 3.42% | 0.48% |

In addition to the significant change in the sensitivity, the performance of each system changes dramatically with the size of the reference. Table 4 shows the performance of each system in reads per second for both versions of the reference. The final column is the percent decrease in performance when the reference is expanded to include alternative splicings. GSNAP is slightly faster than BWA-SW with no alternative splicings, but suffers the greatest performance degradation with the larger reference. The FPGA version is significantly faster than any of the software-based aligners for a single thread compared to a single FPGA. The speedup of 5.57x over BWA with alternative splicings is significant, especially when considering that the results are also of higher quality. The performance of multiple BWA-SW threads compared to multiple FPGAs is easy to extrapolate, since they both scale relatively linearly.

TABLE 4
ALIGNER PERFORMANCE WITH AND WITHOUT AL-
TERNATIVE SPLICINGS

| Aligner | Without Alternative Splicings (reads / sec) | With Alternative Splicings (reads / sec) | Alternative Splicings Performance Reduction |
|---------|---------|---------|---------|
| BFAST | 266 | N/A | N/A |
| GSNAP | 926 | 273 | 70.5% |
| BWASW | 698 | 623 | 10.7% |
| FPGA | 5834 | 3467 | 40.6% |

These results demonstrate a scenario where the flexibility and robustness of our FPGA accelerated solution really shine. In particular, BFAST is completely incapable of handling the alternative splicings and GSNAP suffers a severe degradation in performance. Although the FPGA system appears to suffer more than BWA-SW, it is important to note that no parameter tuning was done for this problem whatsoever, and it is our belief that by tweaking CAL / seed limits and other software parameters, none of which requires rebuilding bitfiles on the V2 system, this performance degradation could be mitigated or even eliminated without decreasing the quality of results. In particular, adding complete alternative splicings (as we did in these benchmarks) – instead of only the exon junctions and enough bases to cover the read length on either side – creates a number of identical regions where the same exon configuration repeats. For example, in the case of 5 exons, the overlap between exon 1 and 2 would exist for splicing 1-2-3, 1-2-4, 1-2-3-5 and so on. RNA-Seq data is likely to hit these regions frequently greatly increasing the number of CALs per read. Each possible splicing produces a separate CAL, which is a situation in which BWT based aligners like BWA-SW are much less susceptible to performance degradation.

In order to address the CALS per read issue raised above the V2 FPGA-based aligner allows for a limit on the maximum number of CALs allowed per seed during index construction. As discussed previously, this is the mode of operation used by BFAST and the V1 system. The V2 system also offers the ability to easily vary the seed length, with shorter seeds reducing performance and improving quality. The results of an experiment exploring these tradeoffs appear in Table 5. Note that the default values that have been used for all previous experiments are seed length of 25 and no CAL limit (the shaded row). These results demonstrate that limiting CALs has a huge impact on performance, resulting in a 12.8x speedup in the case of 25 base seeds, while only having a very minor impact on the quality of results. This improvement comes from processing vastly fewer redundant CALs / read. Conversely, reducing seed length results in a very minor quality improvement at a significant increase in CALs / read and a corresponding decrease in reads / sec. Note that after the 1024 CAL limit, CIGAR string generation becomes a bottleneck for our system. The 128 CAL limit rows do not include faster CIGAR string generation, as this is an area of research in its own right and outside the scope of this work.

TABLE 5
V2 FPGA ALIGNER QUALITY RESULTS (PERFORMANCE NUMBERS FOR HUMAN READS) WITH ALTERNATIVE SPLICINGS USING DIFFERENT INDEX PARAMATERS

| Seed Len | CAL Limit | Correct | Incorrect | Tie | Unaligned | kReads / sec |
|----------|-----------|---------|-----------|-----|-----------|--------------|
| | ∞ | 95.6% | 0.77% | 3.45% | 0.14% | 0.940 |
| 20 | 1024 | 95.4% | 0.88% | 3.44% | 0.26% | 39.7 |
| | 128 | 95.0% | 1.04% | 3.42% | 0.50% | 178 |
| | ∞ | 95.4% | 0.72% | 3.42% | 0.48% | 3.47 |
| 25 | 1024 | 95.1% | 0.73% | 3.42% | 0.73% | 44.4 |
| | 128 | 94.9% | 0.71% | 3.41% | 0.99% | 272 |

Next, we will compare the energy required for the FPGA-based solution and BWA. One way to measure the energy required for a computation is to look at the power consumption of the entire system for the length of that computation. Using this method we calculated the energy required to align one million reads in each system, seen in Table 6. Each row represents an aligner, with the first row being the idle power of the system. The 2 FPGA system results are included to demonstrate the benefit of adding additional FPGAs that can share the software threads feeding them data. Although the FPGA system used the most power, it was also the most energy efficient due to the shorter runtime. Adding more FPGAs increases efficiency even further as expected, and although we only have 2 FPGA boards available for testing this effect should scale with more. The energy comparisons are done with the idle power subtracted out, to be as free from the specifics of our system as possible. If total system power is used, the FPGA advantage becomes much larger. The energy per read, in joules, is simply computed as (Alignment Power – Idle Power) * (seconds / read).

TABLE 6
ADDITIONAL ALIGNER ENERGY COMPARISON

| Aligner | Power (W) | Time (sec / 1M reads) | Energy (J / read) |
|---------|-----------|------------------------|-------------------|
| Idle | 286 | N/A | N/A |
| BWASW | 303 | 1433 | 0.024 |
| 1 FPGA | 350 | 300 | 0.019 |
| 2 FPGAs | 384 | 175 | 0.017 |

Using a single FPGA decreases energy per read by 20.1% compared to BWA-SW and using two FPGAs decreases energy by 29.2% compared to a single BWA thread. Because additional BWA threads do not scale as well as FPGA threads, due to the lack of shared compute resources between them, the advantage actually increases when comparing multiple BWA threads to multiple FPGAs, but that analysis is outside the scope of this work.

Finally, we compare our work to other FPGA accelerated short read aligners (Table 7). We have done our best to compare results fairly, but the limited details of the experiments provided for some systems and the fact that each evaluation uses different datasets, may introduce some additional variation. Our aligner, without alternative splicings, performs similarly to [23] and [24], despite being more accurate than BWA. We believe our higher accuracy comes from using a CAL table based method, as opposed to the FM-index based method of [23] and BWA.

Although [24] uses a hash-based approach, similar to ours, the seeds are masked, as in GSNAP. This may explain the lower accuracy but increased performance for real world data, with potentially many mismatches from the reference (see Fig. 8 in that work). A significantly larger FPGA is also required, and given that our performance scales with PEs, it's not clear which system would perform better on an FPGA of the same size. [25] is different, in that it is very FPGA resource efficient but requires a great deal of offchip DRAM to store expanded FM-indexes. The system described uses 8 FPGAs, with slightly lower per-FPGA performance. This is very difficult to compare to our system, since it is unlikely we could maintain linear scaling up to 8 FPGAs with a single host, due to the memory bandwidth requirements.

TABLE 7
COMPARISON TO FPGA ACCELERATED ALIGNERS, NORMALIZED TO A SINGLE FPGA

| Aligner | Device | kLUTs | BRAMs (Mb) | kReads / sec |
|---------|--------|-------|------------|--------------|
| Our V2 | Virtex-7 | 191 | 5.6 | 336 |
| [23] | Stratix V | 235 | 50.0 | 313 |
| [24] | Virtex-7 | 336 | 31.5 | 511 |
| [25] | Stratix V | 75 | 17.6 | 198 |

# 7 CONCLUSION

In this article, we presented a novel FPGA-based system for short read alignment. This system is a 3.5x improvement over the previously published version and is also 8.4x to 5.6x faster than similar software-based aligners and can be tuned to reach speedups of 71x while maintaining higher quality. The system makes use of the available memory bandwidth on the FPGA board as well as the much greater host memory bandwidth to achieve this high performance while also improving flexibility and configurability to support configurations similar to a wide variety of popular software-based alignment algorithms.

In addition to the improved runtime and flexibility, faster short read alignment that doesn't slow down with large reference genomes can improve the quality of results in many applications. One of these is identifying the source genome for cross-species cell co-cultures or xenografts, where incorrect or missing identification of source genome was reduced by 53% compared to the popular BWA-SW algorithm while also reducing per read energy.

## Acknowledgment

The authors would like to thank Micron Technology, Inc. (formerly Pico Computing) for their support and Daniel C. Jones for generating the mouse and human transcriptomes. The authors are also grateful for the contributions of Corey B. Olson, Maria Kim, Cooper Clauson, Boris Kogon and Carl Ebeling to the original system.

This material is based upon work supported by the National Science Foundation Graduate Research Fellowship under Grant No. DGE-0718124.

WLR was supported in part by NIH 5P01GM081619-08.

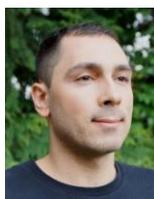

**Nathaniel McVicar** received his BS in Electrical and Computer Engineering from Washington University in St. Louis in 2006, after which he developed FPGA accelerated market data platforms at Exegy, Inc. He received a MS in Electrical Engineering from University of Washington in 2011, where he was an NSF Graduate Research Fellow. His current research interests are reconfigurable computing and bioinformatics accelerators.

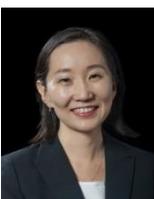

**Akina Hoshino** received her BA in Biology and Chinese from Vassar College and her PhD in Neuroscience from University of Maryland, Baltimore. She is currently a postdoctoral fellow in the lab of Dr. Thomas Reh at the University of Washington. Her research is focused on understanding human fetal retinal development and to use this knowledge to improve retinal differentiation of pluripotent stem cells.

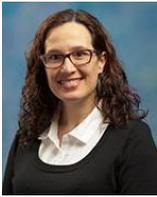

**Anna La Torre** received a BS degree in Biology from the University of Barcelona, and a PhD in Neurobiology from the Institute for Research in Biomedicine Barcelona (IRBB)-University of Barcelona in 2008. She is an Assistant Professor in the University of California Davis. Her research include understanding some of the mechanisms that regulate retinal differentiation as well as the molecular cues that regulate cell acquisition during normal development and in stem cell cultures. She is a member of ARVO (Association for Research in Vision and Ophthalmology), ISER (International Society for Eye Research) and SDB (Society for Developmental Biology).

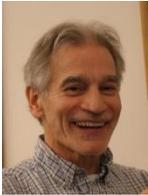

**Thomas A. Reh** has been studying development and regeneration in the retina since 1985. His group developed methods for retinal cell culture and used this system to identify key factors in retinal development and regeneration. The Reh lab was among the first to culture human fetal retina and using what they learned from studying retinal stem/progenitors in vitro over many years, they developed one of the first protocols for directing hES cells to the retinal progenitor cell fate. The lab also carried out pioneering molecular analyses of the developing human retina, using RNAseq and DNase-I hypersensitivity.

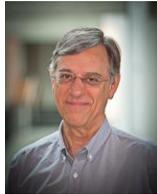

**Walter L. Ruzzo** received a B.S. (in Mathematics) from the California Institute of Technology in 1968, a Ph.D. (Computer Science) from the University of California at Berkeley in 1978, and has been with the University of Washington since 1977 where he is currently a Professor in the Allen School of Computer Science and Engineering, an Adjunct Professor of Genome Sciences and is a Joint Member of the Fred Hutchinson Cancer Research Center. His research is focused on development of computational methods and tools for molecular biology, genetics and genomics.

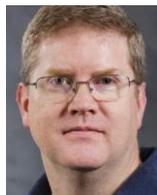

**Scott Hauck** received the BS degree in Electrical Engineering and Computer Science from the University of California - Berkeley in 1990, and the MS and PhD degrees in Computer Science and Engineering from the University of Washington in 1992 and 1995 respectively. He is the Gaetano Borriello Professor for Educational Excellence in the University of Washington's Department of Electrical Engineering. His research focuses on FPGA-based systems, including architectures, applications, and computation tools for these systems. He is an IEEE Fellow, a Sloan Fellow, a recipient of the NSF CAREER award, and multiple best paper and teaching awards.